\documentclass[12pt,reqno,oneside]{amsart}
\usepackage{psfig}

\newcommand{\e}{{\mathrm e}}
\renewcommand{\d}{{\mathrm d}}
\renewcommand{\i}{{\mathrm i}}
\def\lp{\left(}
\def\rp{\right)}
\def\lb{\left[}
\def\rb{\right]}

\def\cF{{\mathcal F}}

\def\cD{{\mathcal D}}

\def\reals{{\mathbb R}}

\def\cmplx{{\mathbb C}}

\def\natnums{{\mathbb N}}

\def\tphi{\tilde{\phi}}

\def\hphi{\hat{\phi}}

\def\halpha{\hat{\alpha}}

\def\homega{\hat{\omega}}

\def\sgn{\mathop{\rm sgn}\nolimits}
\def\arcsinh{\mathop{\rm sinh^{-1}}\nolimits}
\def\arctanh{\mathop{\rm tanh^{-1}}\nolimits}

\newcommand{\expf}[1]{\e^{#1}}

\newtheorem{theorem}{Theorem}

\newtheorem{lemma}[theorem]{Lemma}
\newtheorem{problem}[theorem]{Problem}
\newtheorem*{probnn}{Problem \ref{prob:findpot}A}

\begin{document}
\bibliographystyle{alpha}
\title[The Liouville transformation and exact solutions]{On the Liouville transformation and exactly-solvable Schr\"odinger equations}
\author{Robert Milson\dag}
\thanks{\dag Institute for Advanced Study, Princeton, New Jersey, USA}
\thanks{This research supported by an NSERC, Canada post-doctoral fellowship.}
\address{Robert Milson\\School of Mathematics\\
Institute for Advanced Study\\
Olden Lane\\ Princeton, NJ 08540\\
Telephone: (609) 734 8388 \\
Email: {\tt milson@ias.edu}}
\begin{abstract}
  The present article discusses the connection between
  exactly-solvable Schr\"odinger equations and the Liouville
  transformation.  This transformation yields a large class of
  exactly-solvable potentials, including the exactly-solvable
  potentials introduced by Natanzon.  As well, this class is shown to
  contain two new families of exactly solvable potentials.
\end{abstract}
\maketitle

\section{Introduction}
The study of exactly solvable Schr\"odinger equations dates back to
the very beginnings of quantum mechanics.  As examples one can site
the harmonic oscillator, the Coulomb, the Morse \cite{Morse},
P\"oschll-Teller \cite{PoschlTeller}, Eckart \cite{Eckart}, and the
Manning-Rosen \cite{ManningRosen} potentials.  One can argue that in
each of these cases the exact-solvability comes about because the
Schr\"odinger equations in question can be transformed by a gauge
transformation and by a change of variables into either the
hypergeometric or the confluent hypergeometric equation.  To be more
precise, in each of the above cases there exists a gauge factor
$\sigma(z;E)$, which depends on the energy parameter $E$, and a change
of variables $z=z(r)$, which does not, such that solutions to the
corresponding Schr\"odinger equation,
\begin{equation}
\label{schrod.eqn}
-\psi''(r;E) + U(r)\psi(r;E) = E\psi(r;E),
\end{equation}
are of the form 
\begin{equation}
\label{gaugexform.eqn}
\psi(r;E) = \exp[\sigma(z(r);E)]\,\phi(z(r);E),
\end{equation}
where $\phi(z;E)$ is either $F(\alpha,\beta;\gamma;z)$, the
Gauss hypergeometric function, or $\Phi(\alpha;\gamma;z)$, the
confluent hypergeometric function, and where the parameters $\alpha$,
$\beta$, $\gamma$ are themselves functions of $E$.
The just mentioned types of special functions
are well understood, and as a consequence one can explicitly
calculate the bound state and scattering information for the
corresponding potentials.  In light of these remarks the following
question is of interest.
\begin{problem}
\label{prob:findpot}
Given a collection of functions, $\cF=\{\phi(z)\}$, find all possible
potentials, $U(r)$, such that there exist an $E$-dependent gauge
factor, $\sigma(z;E)$, and an $E$-independent change of variables,
$z(r)$, such that the solutions of equation \eqref{schrod.eqn} are of
the form shown in \eqref{gaugexform.eqn}.
\end{problem}

For the cases of hypergeometric and confluent hypergeometric
functions, Problem \ref{prob:findpot} was solved by Natanzon in
\cite{natanzon1}.  The corresponding classes of exactly-solvable
potentials have come to be known as Natanzon's hypergeometric and
confluent hypergeometric potentials, and have been the subject of some
discussion in the literature \cite{ginocchio} \cite{cordero93}
\cite{wu89}. The purpose of the present article is to review
Natanzon's approach and then to enlarge Natanzon's class of
exactly-solvable potentials by allowing $\phi$ to come from a larger
class of special functions, namely the solutions of the following
class of differential equations:
\begin{equation}
  \label{eq:algform}
  A(z)\phi''(z) + B(z) \phi'(z) + C\phi(z) = 0,
\end{equation}
where $A(z)$ is a non-zero real polynomial of degree 2 or less, $B(z)$
is a real polynomial of degree 1 or less, and $C$ is a real constant.

Prior to Natanzon, Problem \ref{prob:findpot} was considered by Bose
in \cite{bose}, and by several other authors \cite{Manning}
\cite{BhattSud}.  Bose's paper is noteworthy because it introduced the
approach that was followed by Natanzon in his classification.  This
approach relies on two techniques: a certain canonical form for
linear, second-order differential operators, and the Liouville
transformation.  The Liouville transformation will be described in
Section \ref{sect:liouville} and the Bose-Natanzon approach in Section
\ref{sect:natanzon}.  The solution of Problem \ref{prob:findpot} for
the case where $\cF$ is the set of solutions of equation
\eqref{eq:algform} is given in Section \ref{sect:genpot}.  The
resulting collection of potentials includes Natanzon's hypergeometric
and confluent hypergeometric potentials, as well as two new classes of
exactly-solvable potentials.  These new potentials will be discussed
in Section \ref{sect:newpot}.

\section{The Liouville Transformation}
\label{sect:liouville}
Consider a linear, second-order differential equation
\begin{equation}
  \label{eq:sophi}
  a(z)\phi''(z) + b(z)\phi'(z) + c(z)\phi(z) = 0,
\end{equation}
Dividing through by $a(z)$
and making the gauge transformation
\begin{equation}
  \hphi(z) = \exp\lp\int^z \frac{b(t)}{2a(t)} dt\rp\phi(z) 
\end{equation}
changes the equation into the following self-adjoint, canonical form
\begin{equation}
  \label{eq:bosecanform}
  \hphi''(z) + I(z) \hphi(z) = 0, 
\end{equation}
where the potential term is given by
\begin{equation}
  \label{eq:boseinv}
  I = \frac{1}{4a^2}\,(4ac - 2ab'+2ba' - b^2).
\end{equation}
Clearly, $I(z)$ is an invariant of equation \eqref{eq:sophi} with
respect to gauge transformations and multiplication by functions, and
this is why equation \eqref{eq:bosecanform} is being called a
canonical form.  Henceforth, $I(z)$ will be called the Bose invariant
of equation \eqref{eq:sophi}.

A change of the independent variable, say $z=z(r)$, will transform
equation \eqref{eq:bosecanform} into
$$
[z'(r)]^{-2} \tphi''(r) - \frac{z''(r)}{[z'(r)]^3} \tphi'(r) + I(z(r))
\tphi(r) = 0,$$
where $\tphi(r) = \hphi(z(r))$.  The corresponding 
canonical equation is
\begin{equation}
  \label{eq:boseresult}
  \psi''(r) + J(r) \psi(r) = 0.  
\end{equation}
where
\begin{gather}
  \psi(r) = [z'(r)]^{-\frac{1}{2}} \tphi(r),\\
  \label{eq:potrel}
  J(r) = [z'(r)]^2 I(z(r)) + \frac{1}{2} \{ z , r \},  
\end{gather}
and where the curly brackets term denotes the Schwarzian derivative of
$z$ with respect to $r$, namely
$$
\{ z , r \} = \lb\frac{z''(r)}{z'(r)}\rb' - \frac{1}{2} \left[
  \frac{z''(r)}{z'(r)}\right]^2.
$$

The above process of going from one self-adjoint equation to another
by means of a change of variables has been named the Liouville
transformation in \cite{folver}, and the Liouville-Green
transformation in \cite{zwillinger}.  The Liouville transformation
arises naturally in the context of WKB approximation (see Chapter 6 of
\cite{folver}), and also underlies the following classical theorem due
to Schwarz (see \cite{hille}, Theorem 10.1.1 or \cite{polver}, Theorem
6.28 .)
\begin{theorem}
The general solution to the Schwarzian equation
$$\{z,r\} = 2 J(r)$$
has the form $z(r)=\psi_2(r)/\psi_1(r)$, where
$\psi_2(r)$ and $\psi_1(r)$ are two linearly independent, but
otherwise arbitrary solutions of equation \eqref{eq:boseresult}.
\end{theorem}
\noindent
In particular, this theorem implies that for every potential $J(r)$,
there exists a change of variables $z(r)$ such that the corresponding
Liouville transformation takes the equation $\phi''(z) = 0$ to
equation \eqref{eq:boseresult}.  Therefore, one can relate any two
equations of form \eqref{eq:boseresult} by a Liouville transformation.
It is for this reason that Problem \ref{prob:findpot} must be
formulated with the condition that $z(r)$ not depend on the energy
parameter.  Without this restriction the problem would be
uninteresting; one would get a criterion that would be satisfied by
all possible potentials.

\section{The Bose-Natanzon approach}
\label{sect:natanzon}
The approach in question rests on the following reformulation of
Problem \ref{prob:findpot}.
\begin{probnn}
  Given a collection of functions, $\cF=\{\phi(z)\}$, find all
  possible $I_1(z)\geq 0$ and $I_0(z)$ such that for some
  $E$-dependent gauge factor, $\sigma(z;E)$ the solutions of
  \begin{equation}
    \label{eq:schrod1}
    \hphi''(z;E) + \lb I_1(z)E+I_0(z)\rb\hphi(z;E) = 0, 
  \end{equation}
  are of the form
  $$\hphi(z;E) = \exp(\sigma(z;E))\, \phi(z;E).$$
\end{probnn}
Indeed, suppose that $I_1$ and $I_0$ satisfy the above set of requirements.
Let $z(r)$ be a solution of 
\begin{equation}
  \label{eq:xvarrel}
z'(r) = \lp I_1(z)\rp^{-\frac{1}{2}},
\end{equation}
From formula \eqref{eq:potrel} it follows that a Liouville
transformation of equation \eqref{eq:schrod1} based on the change of
variables $z=z(r)$ yields an equation with potential term $E-U(r)$
where
\begin{equation}
  \label{eq:upotform}
  -U(r) = \frac{I_0(z)}{I_1(z)} +
  \frac{-4\,I_1(z)I_1''(z)+5\,(I_1'(z))^2}{16\,(I_1(z))^3}.  
\end{equation}
Furthermore, the corresponding eigenfunctions will have the form
$$
\psi(r;E) = \lp I_1(z) \rp^{\frac{1}{4}}
\,\hphi(z;E).
$$
Therefore $U(r)$ satisfies the criterion imposed by Problem
\ref{prob:findpot}.  One can also reverse the above argument to show
that given a $U(r)$ demanded by Problem \ref{prob:findpot}, one can
produce an $I_1$ and an $I_0$ that satisfy the criterion of Problem
\ref{prob:findpot}A.  In other words, the two formulations are
equivalent.

The Bose invariant for the hypergeometric equation,
$$z(1-z)\phi''(z) + (\gamma-(1+\alpha+\beta))\,\phi'(z) -
\alpha\beta\,\phi(z)=0, $$
is given by
$$I(z) = \frac{T(z)}{4z^2(1-z)^2},$$
where 
$$T(z) = (1-(\alpha-\beta)^2)z^2 + 
(2\gamma(\alpha+\beta-1)-4\alpha\beta)z+\gamma(2-\gamma)
$$
Note that every polynomial, $T(z)$, of degree 2 or less can be
obtained from some choice of $\alpha$, $\beta$, $\gamma$.  Therefore,
in order to solve Problem \ref{prob:findpot}A one must determine all
possible $I_1(z)$ and $I_0(z)$ such that for all $E$ there exists a
$T(z;E)$ of degree two or less in $z$ such that
$$I_1(z)E+I_0(z) = \frac{T(z;E)}{4z^2(1-z)^2}.$$
It is clear that this
condition is satisfied if and only if $T(z;E) = R(z)E + S(z)$ where
$R(z)$ and $S(z)$ are polynomials of degree 2 or less, and such that
$R(z)\geq0$ in the domain of interest.  The determining relation for
$z(r)$ follows from \eqref{eq:xvarrel}; it is
$$
  z'(r) = \frac{2z(1-z)}{\sqrt{R(z)}}.  
$$
Setting $R(z)=r_2z^2+r_1z+r_0$, calculating $\{z,r\}$ and plugging the
result into \eqref{eq:potrel} one obtains the formula for Natanzon's
hypergeometric potentials:
{\smaller[2]
\begin{equation}
  \label{eq:natanzonpot}
  U= \frac{-S(z)+1}{R(z)} +
  \lp \frac{r_1-2(r_2+r_1)z}{z(1-z)} -  \frac{5}{4}\frac{(r_1^2-4
  r_2r_0)}{R(z)} + r_2 \rp \frac{z^2(1-z)^2}{R(z)^2}
\end{equation}}
\noindent
These potentials describe the solution of Problem \ref{prob:findpot}
for the case where $\cF$ is the set of hypergeometric functions.

It is well known that one can transform hypergeometric functions into
confluent hypergeometric ones by a certain limit process.  Natanzon
obtained his confluent hypergeometric potentials by applying this
limit process to his hypergeometric potentials.  The resulting family
of potentials can also be considered as a solution of Problem 1, but
the corresponding $\cF$ is not the set of confluent hypergeometric
functions, but rather the set of {\em scaled} confluent hypergeometric
functions; namely $\phi(z)=\Phi(\alpha;\gamma;\omega z)$, where
$\omega$ is an extra scaling parameter.  These functions satisfy the
following scaled version of the confluent hypergeometric equation
$$z\phi''(z) + (\gamma-\omega z)\,\phi'(z)-\omega\alpha\phi(z)=0.$$
The corresponding Bose invariant is
$$I(z) = \frac{-\omega^2 z^2 +
  2\omega(\gamma-2\alpha)z+\gamma(2-\gamma)}{4z^2},
$$
where again every possible second degree polynomial can occur in the
numerator as one varies $\alpha$, $\gamma$, $\omega$. Hence, by the
same reasoning as above, the criterion of Problem
\ref{prob:findpot}A will be satisfied if and only if 
$$I_1(z)E+I_0(z) = \frac{R(z)E + S(z)}{4z^2},$$
where $R(z)$ and $S(z)$ are polynomials of degree 2 or less,
such that $R(z)\geq0$ in the domain of interest.
The determining relation
for $z(r)$ follows from \eqref{eq:xvarrel}; it is
$$
  z'(r) = \frac{2z}{\sqrt{R(z)}}.  
$$
Setting $R(z)=r_2z^2+r_1z+r_0$, calculating $\{z,r\}$ and plugging the
result into \eqref{eq:potrel} one obtains the formula for Natanzon's
confluent hypergeometric potentials:
\begin{equation}
  \label{eq:natconfpot}
  U= \frac{-S(z)+1}{R(z)} +
  \lp \frac{r_1}{z} -  \frac{5}{4}\frac{(r_1^2-4
  r_2r_0)}{R(z)} - r_2 \rp \frac{z^2}{R(z)^2}
\end{equation}

\section{Generalized Natanzon potentials}
\label{sect:genpot}
The present section is devoted to the solution of Problem
\ref{prob:findpot} for the case where $\cF$ is the set of solutions,
$\phi(z)$, of equations of type \eqref{eq:algform}.  The Bose
invariant for equation \eqref{eq:algform} is given by
$$
  I(z)= \frac{T(z)}{ A(z)^2},
$$
where $T(z)$ is a polynomial of degree 2 or less determined by
quadratic combinations of the coefficients of $A(z)$, $B(z)$, and $C$.
The exact formula for $T(z)$ is not important; it can be readily
recovered from equation \eqref{eq:boseinv}.  What is significant, is
that for a fixed $A(z)$, one can obtain every possible $T(z)$ of
degree 2 or less, from some choice of $B(z)$ and $C$.  Consequently,
using the reformulation given by Problem \ref{prob:findpot}A, one must
seek all possible $I_1(z)$ and $I_0(z)$ such that for all values of
$E$, there exist $T(z;E)$ and $A(z;E)$ both of degree 2 or less, such
that 
\begin{equation}
  \label{eq:tacrit}
  I_1(z)E+I_0(z) = \frac{T(z;E)}{ A(z;E)^2}.  
\end{equation}
\begin{lemma}
  \label{lem:iform}
  Suppose that for all $E$ there exist $T(z;E)$ and $A(z;E)$ such that
  relation \eqref{eq:tacrit} holds.  Then, there exist $E$-independent
  polynomials $A(z)$, $R(z)$, $S(z)$ of degree two or less such that
  $I_1$ = $R/A^2$ and $I_0=S/A^2$.
\end{lemma}
\begin{proof}
  Setting $E=0$ in \eqref{eq:tacrit}, one infers that $I_0=T_0/A_0^2$
  where the degrees of $T_0$ and $A_0$ are two or less.  Similarly by
  setting $E=1$ one infers that $I_1 = T_0/A_0^2 - T_1/A_1^2$ where
  the degrees of $T_1$ and $A_1$ are two or less.  Thus, relation
  \eqref{eq:tacrit} may be rewritten as 
  \begin{equation}
    \label{eq:tacrit1}
    \frac{ET_1}{A_1^2} + \frac{(1-E)T_0}{A_0^2} = 
    \frac{E T_1 A_0^2 + (1-E) T_0 A_1^2}{A_0^2 A_1^2} = 
    \frac{T(z;E)}{A(z;E)^2}.
  \end{equation}
  Given rational functions $P_0/Q_0$ and $P_1/Q_1$ where the
  numerators and denominators are relatively prime polynomials, it's
  easy to show that the denominator of the reduced form of
  $$
  \frac{P_0}{Q_0} + \lambda \frac{P_1}{Q_1},\quad
  \lambda\in\cmplx$$
  is the least common multiple of $Q_0$ and $Q_1$
  for all but a finite number of $\lambda$ values. This observation
  implies that the least common multiple of $A_1$ and $A_0$ must have
  degree less than or equal to $2$.
  
  The rest of the proof will be done by cases, based on the degree of
  $A_1$ and $A_0$.  If both $A_0$ and $A_1$ are constants then there
  is nothing to prove. 
  
  Next, consider the case where one of $A_0$ or $A_1$ is a constant,
  but the other one isn't.  Without loss of generality suppose it is
  $A_1$ that is constant.  Then $T_1$ must be constant also, for
  otherwise the linear combinations of $T_0A_1^2$ and $T_1 A_0^2$ would not
  always yield polynomials of degree 2 or less.  Therefore the lemma
  is true for this case also.
  
  Suppose next that both $A_1$ and $A_0$ have degree 1.  If $A_1$ and
  $A_0$ have the same root, then there is nothing to prove.  If $A_1$
  and $A_0$ have different roots, then both $T_1$ and $T_0$ must be
  constants, because otherwise linear combinations of $T_0A_1^2$ and
  $T_1 A_0^2$ would not always yield polynomials of degree 2 or less.
  Hence one can write 
  $$I_1=\frac{T_1A_0^2}{A_0^2A_1^2},\qquad I_0=
  \frac{T_0A_1^2}{A_0^2A_1^2},
  $$
  and this proves the lemma for the case under consideration.
  
  Finally, suppose that one or both of $A_1$ and $A_0$ is second
  degree.  Without loss of generality assume that $A_1$ has degree 2.
  Consequently, $A_1$ must be the least common multiple of $A_0$ and
  $A_1$, i.e. $A_0$ must be a factor of $A_1$.  Now $A_0$ cannot be a
  constant, because generically, the degree of linear combinations of
  $T_1A_0^2$ and $T_0 A_1^2$ would be greater than $2$.  If the degree
  of $A_0$ is $2$, then there is nothing to prove.  The last
  possibility is that $A_1=\Delta A_0$, where both $A_0$ and $\Delta$
  have degree $1$.  In this case
  $$I_1 E + I_0 = \frac{E T_1+(1-E)\Delta^2 T_0}{A_1^2},$$
  and  hence $T_0$ must be a constant.  But one can therefore write
  $$I_0 = \frac{T_0\Delta^2}{A_1^2},$$
  i.e. the lemma is also true for this last case.
\end{proof}

Using the above Lemma, as well as the formulas \eqref{eq:xvarrel} and
\eqref{eq:upotform} one can now give the solution of Problem
\ref{prob:findpot} for the case where $\cF$ is the set of solutions to
equation \eqref{eq:algform}.  The desired potentials have the form
{\smaller[2]
\begin{equation}
  \label{eq:genpot}
  U(r) = \frac{-S(z)+\cD(A)/4}{R(z)} + 
  \lp -\frac{3 R''(z)}{2} - \frac{5}{4}\frac{\cD(R)}{R(z)} + 
  \frac{R'(z) A'(z)}{A(z)}\rp \, \frac{A(z)^2}{4R(z)^2},
\end{equation}
}
where $A(z)$, $R(z)$, $S(z)$ are polynomials of degree two or less,
where $\cD$ denotes the discriminant operator, and where $z(r)$ is a
solution of
\begin{equation}
  \label{eq:genxvar}
  z'(r) = \frac{A(z)}{\sqrt{R(z)}}.  
\end{equation}
Henceforth this class of
potentials will be referred to as the generalized Natanzon potentials.

\section{New exactly-solvable potentials}
\label{sect:newpot}
Note that the presentation of the generalized Natanzon potentials
given by equations \eqref{eq:genpot} and \eqref{eq:genxvar} is
invariant under affine substitutions, $z\mapsto az+b$.  Consequently,
no generality will be lost if one restricts $A(z)$ to one of the
following $5$ possibilities: 1, $z$, $z^2$, $z(z-1)$, $z^2+1$.  The
second and the fourth case yield, respectively, the Natanzon confluent
hypergeometric, and the Natanzon hypergeometric potentials.  If
$A(z)=z^2$, then the substitution $z\mapsto 1/z$ transforms
\eqref{eq:genpot} and \eqref{eq:genxvar} into the corresponding forms
for the case of $A(z)=z$.  Therefore, if $A(z)=z^2$, one again obtains
the Natanzon confluent hypergeometric potentials.  With a bit of work
one can check that at the level of solutions to the respective forms
of equation \eqref{eq:algform}, this transformation corresponds to the
to the well known identity (see chapter 6.6 of \cite{bateman}):
{\smaller[1]
$$
{}_2F_0(\alpha,\alpha+1-\gamma; 1/z) = z^\alpha
\Psi(\alpha,\gamma;z).
$$}
where $\Psi$ is the confluent hypergeometric
function of the second kind.

The two remaining cases, $A(z)=1$ and $A(z)=1+z^2$, yield new families
of exactly solvable potentials. These will be referred to,
respectively, as case 1, and case 5 potentials, and will now be
examined in some detail.  In each case it will be convenient to
rewrite equation \eqref{eq:algform} using a certain choice of adapted
parameters.  These adapted parameters will be denoted by Greek
letters, and the resulting equation will be referred to as the primary
equation.

The solutions of the primary equations can be given in terms of
hypergeometric functions, and have a natural dependence on the adapted
parameters.  The actual potential depends on a choice of $R(z)$.  The
potential parameters --- these will be denoted using lower case Latin
letters --- and the energy parameter, $E$, will turn out to be related
to the adapted parameters by polynomial relations.  It is important to
note that for fixed potential parameters, and a fixed value of $E$ one
must solve these relations in order to obtain the corresponding values
of the adapted parameters of the primary equation.

An examination of formula \eqref{eq:genpot} shows that in order for
$U(r)$ to be non-singular, the $z$-domain of the function shown in the
right hand side of \eqref{eq:genpot} must not contain any roots of
$R(z)$.  Singular potentials will not be discussed here, and this
constraint greatly reduces the possible choices for $R(z)$.

According to equation \eqref{eq:genxvar} the physical coordinate of
the corresponding Natanzon potential is given by
$$r(z)=\int^z \frac{\sqrt{R(t)}}{A(t)}\, dt.$$
In all cases one can
explicitly calculate the above anti-derivative, but the inverse, i.e.
$z(r)$, cannot in general be specified explicitly.  Indeed, one may
reasonably speculate that historically the study of Natanzon
potentials was delayed by the fact that the inverse, $z(r)$, of the
above anti-derivative can be given in terms of elementary functions
only for certain restricted choices of $R(z)$ and $A(z)$ (based on the
remarks found in the first paragraph of \cite{natanzon1} it would seem
that Natanzon shares this viewpoint.)  

Nonetheless, a great deal of information about the inverse is
available.  First, one can always take the domain of $z(r)$ to be the
whole real line; this is a consequence of the fact that one excluded
those $R(z)$ that have roots in the domain of $z$.  One can calculate
power series and asymptotic expansions for $z(r)$ and use these as the
basis for a numerical approximation.  The graphs of potential curves
that are given below were generated using this approach.

In the subsequent discussion $\Phi$ and $\Psi$ will denote the
confluent hypergeometric functions of the first and second kind, and
$F$ will denote the usual hypergeometric function, ${}_2F_1$.  For the
source of this notation, as well as the various properties of these
functions the reader is referred to \cite{bateman}.

\noindent{\em Case 1.}  $A(z) = 1$.
\par\noindent{\em Primary equation: }
$
\phi''(z) - 2\omega(z-\beta)\phi'(z)-4\alpha\omega\,\phi(z) = 0
$
\par\noindent{\em Primary solutions: }
$$
\phi_0=\Phi\!\lp\alpha;\;1/2;\;\omega(z-\beta)^2\rp,\quad
\phi_1 = (z-\beta)\,\Phi\!\lp\alpha+1/2;\;3/2;\;\omega (z-\beta)^2\rp\
$$
Note that the family of potentials under discussion is invariant under
substitutions of the form $z\mapsto z+k$, $k\in\reals$.
In order to obtain a non-singular potential, $R(z)$ must not have real
roots, and consequently
it is sufficient to consider the case $R(z)=z^2+1$.  The most general
form of the potential can then be obtained by a scaling transformation.

\par\noindent{\em Distance 1-form and physical variable:}
\begin{align}
\nonumber
\d r &=\sqrt{z^2+1}\,\d z, \qquad
r=\frac{1}{2} \lp z\sqrt{\strut z^2+1} + \log(z+\sqrt{z^2+1})\rp.\\
\label{case1-zr.eqn}
z &\sim \pm \sqrt{2|r|},\quad r\rightarrow  \pm\infty .
\end{align}

\begin{figure}[p]
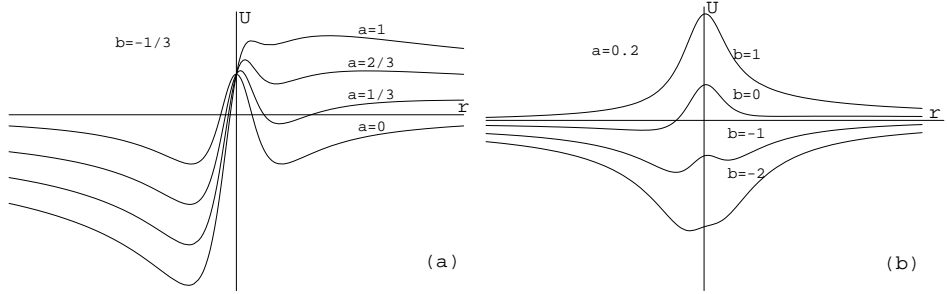

\centering\noindent
\psfig{figure=case1-a.epsi,width=2.4in}
\hfil
\psfig{figure=case1-b.epsi,width=2.4in}
\caption{Case 1: roles of the $a$ and $b$ parameters.}
\label{case1.fig}
\end{figure}

\noindent{\em Potential:}
\begin{align}
\nonumber
U&=\frac{az+b}{z^2+1} -\frac{3/4}{(z^2+1)^2}+\frac{ 5/4}{(z^2+1)^3} \\
\label{case1-parms.eqn}
E& = -\omega^2,\quad
a=-2\beta\omega^2,\quad
b=\omega^2(\beta^2-1)-\omega(1-4\alpha).
\end{align}
The resulting 2-parameter family of potentials are characterized by
the presence of two wells separated by a barrier --- see Figure
\ref{case1.fig}.  The parameter $a$ 
controls the degree of asymmetry; if $a=0$, the potential is symmetric about
$r=0$.  The parameter $b$ controls the height of the central barrier.
As $b$ increases the wells become smaller; if in addition $a=0$, then they
disappear altogether.  As $b$ decreases the  two wells merge into one.
\par\noindent{\em Eigenfunctions:}
\begin{equation}
\label{case1-efuncs.eqn}
\psi_i = \expf{-1/2\omega(z-\beta)^2} (z^2+1)^\frac{1}{4} \phi_i,\;
i=0,1.
\end{equation}

\noindent{\em Bound states:} 

\noindent
A bound state occurs when $-2\alpha\in\natnums$, and when
$\omega>0$.  This directly implies that there are infinitely many bound states
if $a\neq 0$ or if $b<0$; and that there are no bound states, otherwise.

\noindent{\em Scattering.}

\noindent
The scattering states occur when $E>0$.  Correspondingly,
$\omega=i\homega$, and $\alpha=\frac{1}{4}+\i\halpha$, where $\homega>0$ and
$\halpha$ are real.  As a consequence $\psi_0$ and
$\psi_1$ are real valued functions; this follows directly from 
the well known formula (see chapter 6.3 of \cite{bateman}):
$$
\Phi(a;c;z) = \e^z\Phi(c-a;c;-z).
$$
An asymptotically free eigenfunction, call it $\psi_{\rm f}$,  can be given
explicitly in terms of the 
confluent hypergeometric function of the second kind:
$$\psi_{\rm f}=\omega^\alpha \expf{-1/2\omega(z-\beta)^2} (z^2+1)^\frac{1}{4}
\Psi(\alpha;1/2;\omega(z-\beta)^2).$$
From the well-known asymptotic formula (see chapter 6.13 of \cite{bateman})
$$
\Psi(a;c;z) \sim z^{-a} ,\quad z\rightarrow +\infty,
$$
and from (\ref{case1-zr.eqn}) it follows that
$$
\psi_{\rm f} \sim \expf{-\i\homega |r|} 
\expf{\i\lp\pm \frac{a}{\homega^2\sqrt2}\sqrt{|r|} -\frac{ a^2}{4\homega^3}\rp}
\lp\sqrt{2|r|}\mp\beta\rp^{2\i\halpha},
$$
as $r\rightarrow \pm\infty$.  Thus, asymptotically $\psi_{\rm f}$
represents an almost free particle traveling toward the center.  The
discontinuity in the direction of motion is caused by the fact that
$\Psi(z)$ is not regular at $z=0$.  The extra terms in the asymptotic
phase appear because of the slow rate --- on the order of $r^{-1}$ an
$r^{-1/2}$ , depending respectively on whether $a$ is zero or not ---
at which the potential falls off toward zero.

From the relation between confluent hypergeometric functions of the
first and second kind (see chapter 6.7 of \cite{bateman}) one obtains
$$
\psi_0 = c_0 \psi_{\rm f} + \overline{c_0 \psi_{\rm f}},\quad
\psi_1 = \sgn(z)(c_1 \psi_{\rm f} + \overline{c_1 \psi_{\rm f}}).
$$
where
$$c_0 = (-\omega)^{-\alpha}
\frac{\Gamma(\frac{1}{2})}{\Gamma(\frac{1}{4}-\i\halpha)},\quad 
c_1 =  (-\omega)^{-\frac{1}{2}-\alpha}
\frac{\Gamma(\frac{3}{2})}{\Gamma(\frac{3}{4}-\i\halpha)}
$$
It immediately follows that
$$\frac{1}{2}\lp\frac{\psi_0}{ c_0} - \frac{\psi_1}{ c_1}\rp =
\begin{cases}
T \bar{\psi_{\rm f}}, & z\rightarrow +\infty \\
\psi_{\rm f} + R\bar{\psi_{\rm f}}\; &  z\rightarrow -\infty
\end{cases}
$$
where the reflection and transmission coefficients are given by
$$
T = \expf{\i\theta}(1-\expf{-2\alpha\pi \i})^{-1},\quad
R = \expf{\i\theta}(1-\expf{2\alpha\pi \i})^{-1},
$$
where
$$
\expf{\i\theta}
=\frac{\Gamma(\frac{1}{2}-\alpha)}{\Gamma(\alpha)}
E^{(\alpha-\frac{1}{4})}\expf{-\frac{\pi \i}{4}}.
$$

\noindent{\em Case 5.}
$A(z)= z^2+1$
\par\noindent{\em Primary equation: }
$$(1+z^2)\phi''(z) +
(\i(1-2\rho)+(2\sigma+1)z)\phi'(z)+(\sigma^2-\delta^2)\phi=0.$$

\noindent
The above equation is related to the usual hypergeometric equation,
$$\zeta(1-\zeta)\phi_{\zeta\zeta} +
(\gamma-\zeta(\alpha+\beta+1))\phi_\zeta - \alpha\beta\phi = 0,$$
by a linear change of parameters, and a complex-linear change of coordinates:
$$\sigma=\frac{\alpha+\beta}{2},\quad
\delta=\frac{\alpha-\beta}{2},\quad
\rho = \gamma-\frac{\alpha+\beta}{2},\quad
\zeta=\frac{1-\i z}{2}.
$$
\par\noindent{\em Primary solutions: }
\begin{align*}
\phi_1 &= F\lp\sigma+\delta,\, \sigma-\delta;\, \
{\rho+\sigma};\, (1-\i z)/2\rp,\\
\phi_2 &= F\lp\sigma+\delta,\, \sigma-\delta;\, \
{1-\rho+\sigma};\, (1+\i z)/2\rp
\end{align*}

To obtain non-singular potentials one must take $R(z)$ without any
real roots.  The resulting family of potential depends of 4
parameters.  A treatment of the most general potential, i.e. one
depending all 4 parameters would be unduly long, and not particularly
illuminating.  Thus, the focus here will on a more manageable 3
parameter subclass, namely the potentials that correspond to
$R(z)=z^2+a^2$.
\par\noindent{\em Distance 1-form and physical variable:}
\begin{align*}
\d r&=\frac{\sqrt{z^2+a^2}}{ z^2+1}\,\d z,\qquad
r=\arcsinh\lp \frac{z}{a}\rp+
\sqrt{1-a^2} \arctanh\lp \frac{\sqrt{1-a^2} z}{\sqrt{z^2+a^2}}\rp\\
z&\sim
a \exp\lp \sqrt{1-a^2}\,\arctanh\lp\sqrt{1-a^2}\rp\,\rp\,\sinh(r) ,\quad r\rightarrow \pm\infty.
\end{align*}
\par\noindent{\em Potential: }
{\smaller[2]
\begin{align}
  \label{case5-pot.eqn}
  U&=\frac{b+cz}{ a^2+z^2} +\frac{1}{4}\frac{z^2+1}{ z^2+a^2}+
  \lp\frac{z^2+1}{ z^2+a^2}\rp^2
  \lp -\frac{1}{4}-\frac{1}{2}\frac{a^2+1}{ z^2+1}+\frac{5}{4}\frac{a^2}{
    z^2+a^2}\rp \\
  \label{case5-parms.eqn}
  E&=-\delta^2,\; b= \delta^2(1-a^2)-\lp\rho-\frac{1}{2}\rp^2 -
  \lp\sigma-\frac{1}{2}\rp^2 ,\\
  \nonumber c&=-2\i\lp\rho-\frac{1}{2}\rp\lp\sigma-\frac{1}{2}\rp.
\end{align}}
These potentials fall off exponentially toward zero for large  $r$.
Setting $\rho=\frac{1}{2}$ one obtains potentials that coincide with a certain
subclass of  Natanzon hypergeometric potentials.  The  correspondence is
given by the following substitution:
$z\mapsto 1+z^2,$
and at the level of solutions to the respective primary equations is
described by the following quadratic transformation of the
hypergeometric function (see chapter 2.11 of \cite{bateman}):
{\smaller[2]
$$F\lp\sigma+\delta,\,\sigma-\delta;\, 1/2+\sigma; \,(1-\i z)/2\rp 
= F\lp (\sigma+\delta)/2, \,(\sigma-\delta)/2;\,
1/2+\sigma;\,1+z^2\rp.
$$} The generic shape is that of two spikes for $a>1$, and two wells
for $0<a<1$; when $a=1$ one recovers a modified P\"oschll-Teller
potential.  The parameter $b$ controls the height/depth of the central
spike/well, while $c$ is the skew parameter that controls the degree
of asymmetry in the potential.

\begin{figure}[p]
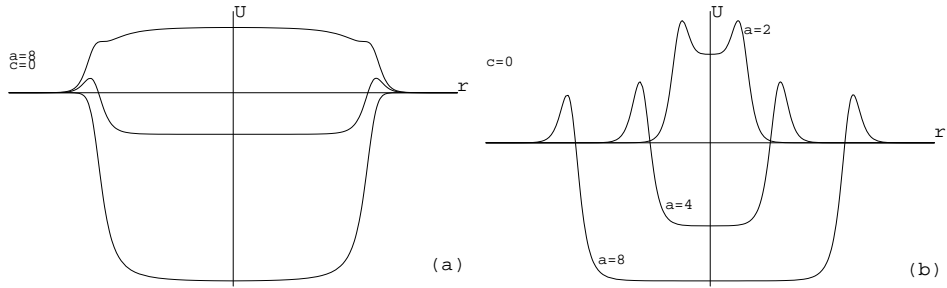

\centering
\noindent
\psfig{figure=case5-1a.epsi,width=2.4in}
\hfil
\psfig{figure=case5-1b.epsi,width=2.4in}
\caption{Case 5 symmetric potentials: (a) critical values of the $b$
  parameter, (b) variation of the $a$ parameter in potentials with the
  middle critical $b$ value.}
\label{case5-1.fig}
\end{figure}

The case $a>1$ results in the more interesting potential shapes,
and thus will be the focus of the remaining discussion.
Consider the symmetric potentials ($c=0$) for a fixed value of
$a>1$.  The number of extrema in the potential curve depends on the
value of $b$.  There are 3 critical values of $b$ where the
number of extrema changes:
$$
\frac{7a^2-3+3a^{-2}}{20},\quad
\frac{-a^2+9-9a^{-2}}{4},\quad
-a^2+\frac{3}{4}.
$$
At these critical values of $b$ some of the extrema merge, and one
obtains some distinguished potential shapes;  these shapes are shown in
Figure \ref{case5-1.fig} (a).

At the value $b=(-a^2+9-9a^{-2})/4$ the potential takes the form
$$
\frac{a^2-1}{ 4a^4}\lp \frac{z^4((a^2-7)z^2+6a^4-12a^2)}{
(z^2+a^2)^3}-(a^2-7)\rp$$
One can show that $z=r/a+O(r^3)$ near $r=0$, and hence the first
3 $r$-derivatives of the potential vanish.  As a
consequence, one obtains a
well with a very flat bottom.  The potential also possesses two
local maxima;  these correspond
to spike-like barriers on either side of the well.  The variation of $a$ in
this type of
potential shape is shown in 
Figure \ref{case5-1.fig} (b).  The two barriers vanish
precisely at the critical value of $b=-a^2+\frac{3}{4}$.

\begin{figure}[p]
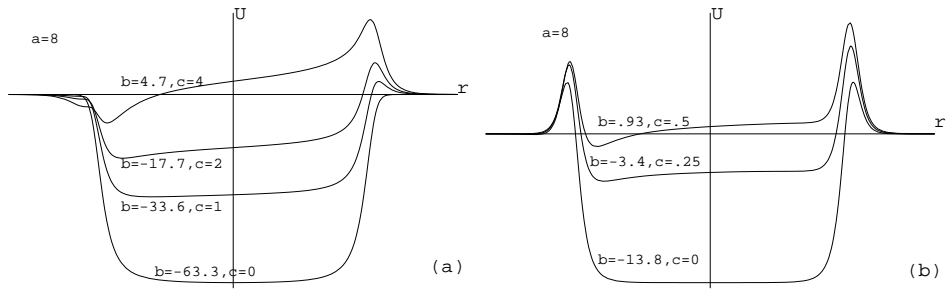

\centering
\noindent
\psfig{figure=case5-2a.epsi,width=2.4in}
\hfil
\psfig{figure=case5-2b.epsi,width=2.4in}
\caption{Case 5 asymmetric potentials: variation of the $c$ parameter
  in potentials with (a) lowest critical $b$ value (b) middle critical
  $b$ value.}
\label{case5-2.fig}
\end{figure}

For $c\neq0$ one can obtain similarly distinguished
potentials whenever $b$ attains a  critical value where
the potential extrema merge.
There is no exact formula for these critical
values of $b$; they must be solved for numerically.  The resulting
asymmetric potentials and their symmetric counterparts are shown
in  Figure \ref{case5-2.fig}.
\par\noindent{\em Eigenfunctions:}
\begin{equation}
\label{case5-efuncs.eqn}
\psi_i =
(1-\i z)^{\frac{\sigma}{2}+\frac{\rho}{2}-\frac{1}{4}}
(1+\i z)^{\frac{\sigma}{2}-\frac{\rho}{2}+\frac{1}{4}}
\lp\frac{a^2+z^2}{ 1+z^2}\rp^\frac{1}{4}\phi_i,\; i=1,2.
\end{equation}
\par\noindent{\em Bound states:} 

\noindent
An examination of relations (\ref{case5-parms.eqn}) will show that in
order to obtain a potential with real coefficients, $\delta$ must
either be real or imaginary; and either $\sigma-\frac{1}{2}$ or
$\rho-\frac{1}{2}$ must be real, while the other must be imaginary.
Note that the following transformation of the parameters is a symmetry of the
potential:
$$
\rho \mapsto \i\lp\sigma-1/2\rp+1/2,\quad
\sigma \mapsto -\i\lp\rho-1/2\rp+1/2.
$$
The transformation $\delta\rightarrow -\delta$ is also a potential
symmetry.  The presence of these two symmetries means that without
loss of generality one can assume that $\sigma$ is real, that
$\Re(\rho)=1/2$, and that $\delta$ is either positive, or
positive imaginary.  With these assumption in place, $\psi_2$ is the
complex conjugate of $\psi_1$, and the latter is square integrable if
and only if $\delta>0$ and $\sigma+\delta\in-\natnums$.  After a bit
of calculation one can show that this criterion implies that for fixed
$a$, $b$, $c$, the bound states are indexed by natural numbers,
$N=-(\sigma+\delta)$ such that
$$N < \frac{\sqrt{-2b+2\sqrt{b^2+c^2}}-1}{2}.$$
In particular, if $c=0$ (the symmetric potentials) then there will be
no bound states if 
$b\geq-1/4$.
If $b<-1/4$, then the number of bound states is equal to the largest
integer smaller than $1/2+\sqrt{-b}$.

\noindent{\em Scattering.}

\noindent
The scattering states occur when $E>0$, and hence without loss of generality
$\delta$ is positive imaginary. 
For reasons detailed above, $\sigma$ will be assumed to be real, while
$\Re(\rho)$ will be assumed to be $1/2$.
To compute the reflection and
transmission coefficients it will be useful to introduce two more
solutions of the primary equation:
\begin{eqnarray*}
\phi_3 &=& \lp\,(\i z-1)/2\,\rp^{-\sigma-\delta}
F\lp\sigma+\delta,\delta+1-\rho;1+2\delta;\, 2/(1-\i z)\rp,\\
\phi_4 &=& \lp\,(\i z-1)/2\,\rp^{-\sigma-\delta}
F\lp\sigma-\delta,\rho-\delta;1-2\delta;\, 2/(1+\i z)\rp.\\
\end{eqnarray*}
As per the formula in (\ref{case5-efuncs.eqn}), let $\psi_3$ and
$\psi_4$ denote the corresponding eigenfunctions.  The usefulness of
$\psi_3$ and $\psi_4$ is that they represent asymptotically free
particles traveling, respectively, towards and away from the origin:
\begin{align*}
\psi_3 &\sim K^{-\delta}
\expf{\mp (\sigma+\rho+\delta-1/2) \, \pi \i/2} \expf{-\delta|r|},\quad&
r\rightarrow \pm \infty,\\
\psi_4 &\sim K^{\delta}
\expf{\mp (\sigma+\rho-\delta-1/2)\, \pi \i/2} \expf{\delta|r|},\quad &
r\rightarrow \pm \infty,\\
\end{align*}
where
$$
K = \frac{a}{4} \exp\lp\, \sqrt{1-a^2}\arctanh\lp\sqrt{1-a^2}\rp\,\rp.
$$
Relations between the regular eigenfunctions, $\psi_1$, $\psi_2$, and
the irregular ones, $\psi_3$, $\psi_4$ are given by (see chapter 2.9
of \cite{bateman}): 
$$
\psi_1 = c_3 \psi_3 + c_4\psi_4,\qquad
\psi_2 = \overline{c_4} \expf{\pm\pi \i (\sigma+\delta)} \psi_3 +
\overline{c_3} \expf{\pm \pi \i(\sigma-\delta)} \psi_4,
$$
where the $\pm$ in the second equation corresponds to the sign of $z$,
and where
$$
c_3 = \frac{\Gamma(\sigma+\rho)\Gamma(2\delta)}{ 
       \Gamma(\rho+\delta)\Gamma(\sigma+\delta)},\quad
c_4 = \frac{\Gamma(\sigma+\rho)\Gamma(-2\delta)}{ 
       \Gamma(\rho-\delta)\Gamma(\sigma-\delta)}.
$$
It follows that
\begin{gather*}
K^\delta \expf{-(\sigma+\delta+\rho-1/2)\, \pi \i/2}
\lp\frac{\psi_1 }{ c_3} \expf{\pi \i(\sigma+\delta)} - \frac{\psi_2
  }{\overline{c_4}} 
\rp = \\
\begin{cases}
T \expf{\delta r}, & r\rightarrow +\infty\\
\expf{\delta r} + R \expf{-\delta r},\; & r\rightarrow -\infty\\
\end{cases}
\end{gather*}
where elementary calculations will show that
\begin{eqnarray*}
T &=& \frac{K^{2\delta}
\Gamma(\sigma-\delta)\Gamma(1-\sigma-\delta)
 \Gamma(\rho-\delta)\Gamma(1-\rho-\delta)}{2\pi\Gamma(-2\delta)
 \Gamma(1-2\delta)} \\ 
R &=& T \lp \frac{\sin(\pi\sigma)\sin(\pi\rho)}{\sin(\pi\delta)}
- \frac{\i\cos(\pi\sigma)\cos(\pi\rho)}{\cos(\pi\delta)} \rp
\end{eqnarray*}

\bibliography{paper4}
\end{document}